\documentstyle[12pt]{article}
\begin{document}
\begin{center}
Ochanomizu University\\
June 14, 2001
\end{center}
\begin{center}
\vspace*{.5cm}
{\Large\bf{Spinors and Supersymmetry}}\vspace{.7cm}\\
{\large{D.G.C. McKeon}}$^\ast$\\
{\large{T.N. Sherry}}$^\dagger$\vspace{.7cm}\\
$^\ast${\small\it{University of Western Ontario}}\vspace{.01cm}\\
$^\dagger${\small\it{National University of Ireland, Galway}}\\
\end{center}
{\small{\bf{Abstract.$\;$}} In this paper, we survey the nature
of spinors and supersymmetry (SUSY) in various types of spaces.
We treat two distinct types of spaces: flat spaces and spaces of constant
(non-zero) curvature.  The flat spaces we consider are either three or
four dimensional of signatures 3 + 1, 4 + 0, 2 + 2 and 3 + 0. In
each of these cases, SUSY generators anti-commute to yield the generators
of translations in the non-compact flat spaces. The spaces of constant
curvature we consider are two-dimensional: the surface of the sphere
$S_2$ and the Anti-deSitter space $AdS_2$. $S_2$ is embedded in a 3 + 0
Euclidean space while $AdS_2$ is embedded in 2 + 1 Minkowski space.
The SUSY generators in these cases anti-commute to yield the generators 
of the isometry groups ($SO(3)$ or $SO(2,1)$) of the space involved.

We also report on some recent developments in looking for superspace
realizations of these SUSY algebras. We can report good progress 
in the $3 + 0$ Euclidean and in the $AdS_2$ case, somewhat less in
the $S_2$ case. In each of the compact cases, we can construct field
multiplet models carrying invariance under the full SUSY algebra.

\section{Flat Space SUSY Analysis}
\subsection{3 + 1 Dimensions}
The analysis of spinors and supersymmetry (SUSY) in three 
dimensional Minkowski space is quite standard.
(See for example ref. [1].)  In a representation in which Dirac matrices
are given by
$$\gamma^\mu = \left[\left(\begin{array}{cc}
0 & 1\\
1 & 0\end{array}\right)\, ,\left(\begin{array}{cc}
0 & \vec{\sigma}\\
-\vec{\sigma} & 0 \end{array}\right)\right]\eqno(1)$$
and a charge conjugation matrix $C$ is defined by
$$C^{-1}\gamma^\mu C = -\gamma^{\mu T},\eqno(2)$$
we have spinors
$$\Psi = \left( \begin{array}{c}
\psi_\alpha\\
\overline{\chi}^{\dot{\alpha}}\end{array}\right)\;\;\;
\overline{\Psi} \equiv \Psi^\dagger \gamma^0 = \left(\chi^\alpha ,
\overline{\psi}_{\dot{\alpha}}\right)\eqno(3a)$$
and
$$\Psi_C \equiv C\overline{\Psi}^T = \left( \begin{array}{c}
\chi_\alpha\\
\overline{\psi}^{\dot{\alpha}}
\end{array}\right)\;\;\;
\overline{\Psi}_C = \left(\psi^\alpha ,
\overline{\chi}_{\dot{\alpha}}\right)\eqno(3b)$$
forming representations of the Lorentz group. The 
spinorial generator $Q$ of the $N = 1$ extension of the Poincar\'{e} group
is Majorana (ie, $Q = Q_C = \displaystyle{\left(\begin{array}{c}
Q_\alpha\\
\overline{Q}^{\dot{\alpha}}\end{array}\right)}$) and
satisfies the algebra
$$\left\lbrace Q_\alpha , Q_\beta \right\rbrace = 0\eqno(4a)$$
$$\left\lbrace Q_\alpha , \overline{Q}_{\dot{\beta}} \right\rbrace = 
2\sigma^\mu_{\alpha{\dot{\beta}}} P_\mu.\eqno(4b)$$
The two spinorial
generators of $N = 2$ SUSY extension of the Poincar\'{e} group are both Majorana
(ie, $Q_i  = Q_{Ci}$ for $i = 1, 2$) and satisfy the algebra
$$\left\lbrace Q_\alpha^i , \overline{Q}_{\dot{\beta}j}\right\rbrace = 2 \delta^i_j
\sigma_{\alpha\dot{\beta}} P_\mu\eqno(5a)$$
$$\left\lbrace Q_\alpha^i , Q_\beta^j\right\rbrace = \epsilon_{\alpha\beta}
\epsilon^{ij} Z\eqno(5b)$$
incorporating a central charge $Z$ which commutes with all the other
generators of this algebra.
A representation of this algebra can be found using Fermionic creation
and annihilation operators
$$a_\alpha = \frac{1}{\sqrt{2}} \left(Q_\alpha^1 +
\epsilon_{\alpha\beta} Q^{2\dagger}_{\;\;\beta}\right),\;\;\;
b_\alpha = \frac{1}{\sqrt{2}} \left(Q_\alpha^1 -
\epsilon_{\alpha\beta} Q^{2\dagger}_{\;\beta}\right).\eqno(6)$$
From (5) it follows that
$$\left\lbrace a_\alpha , a_\beta^\dagger \right\rbrace =
\delta_{\alpha\beta} (2M + Z)\eqno(7a)$$
$$\left\lbrace b_\alpha , b_\beta^\dagger \right\rbrace =
\delta_{\alpha\beta} (2M - Z)\eqno(7b)$$
in a frame in which $P_\mu = (M, \vec{0})$. By (7b) we obtain the
``BPS'' bound
$$2M \geq Z.\eqno(8)$$
\subsection{4 + 0 Dimensions}
The situation in four dimensional Euclidean space is quite different.  In this case
$$\gamma^\mu = \left[\left(\begin{array}{cc}
0 & 1\\
1 & 0\end{array}\right),\;\;\;
\left(\begin{array}{cc}
0 & i\vec{\sigma}\\
-i\vec{\sigma} & 0\end{array}\right)\right]\eqno(9)$$
so that
$$\Psi = \left(\begin{array}{c}
\psi_\alpha\\
\chi^{\dot{\alpha}}\end{array}\right)\;\;\;\;
\overline{\Psi} = \Psi^\dagger = \left(
\overline{\psi}^\alpha\; ,
-\overline{\chi}_{\dot{\alpha}}\right)\eqno(10a)$$
and
$$\Psi_C = C\overline{\Psi}^T = \left(\begin{array}{c}
-\overline{\psi}_\alpha\\
\overline{\chi}^{\dot{\alpha}}\end{array}\right)\;\;\;\;
\overline{\Psi}_C = \left(
\psi^\alpha\; ,
\chi_{\dot{\alpha}}\right)\eqno(10b)$$
for representations of $SO(4) = SU(2) \times SU(2)$. 
The spinor $\psi_\alpha$ transforms under one SU(2) subgroup
while $\chi^{\dot{\alpha}}$ transforms under the other
SU(2) subgroup.
It is also evident from (10) that as
$\left(\overline{\Psi}_C\right)_C = -\Psi$, we cannot have Majorana
spinors in $4 + 0$ dimensions. The simplest self conjugate SUSY algebra is now [2]
$$\left\lbrace Q_a, \overline{R}_{\dot{b}} \right\rbrace = 
i\sigma_{a\dot{b}}^\mu P_\mu\;\;\;\;\left\lbrace Q_a, R_{\dot{b}} \right\rbrace =
0\eqno(11a,b)$$
$$\left\lbrace Q_a, \overline{Q}_b \right\rbrace = \epsilon_{ab} Z_{Q\overline{Q}}\;\;\;\;
\left\lbrace R^{\dot{a}}, \overline{R}^{\dot{b}} \right\rbrace
= \epsilon^{\dot{a}\dot{b}} Z^{R\overline{R}}\eqno(11c,d)$$
or, equivalently, if we define
$$G =  \left(\begin{array}{c}
Q_a\\
R^{\dot{a}}\end{array}\right) \Longrightarrow
\begin{array}{ll}
S_{a1} = Q_a & S_{a2} = \overline{Q}_a\\
T_{\dot{a}1} = -R_{\dot{a}} & T_{\dot{a}2} =
\overline{R}_{\dot{a}}\end{array}
\eqno(12)$$
we obtain an equivalent algebra which displays an $SU(2)$ structure
$$\left\lbrace S_{ai} , S_{bj} \right\rbrace = \epsilon_{ab}
\epsilon_{ij} Z_{Q\overline{Q}}\eqno(13a)$$
$$\left\lbrace T_{\dot{a}i} , T_{\dot{b}j} \right\rbrace = \epsilon_{\dot{a}\dot{b}}
\epsilon_{ij} Z^{R\overline{R}}\eqno(13b)$$
$$\left\lbrace S_{ai} , T_{\dot{b}j} \right\rbrace = i\epsilon_{ij}
\sigma_{a\dot{b}}^\mu P^\mu .\eqno(13c)$$
This is similar in form to (5) with the roles of $Z$ and $P^\mu$
``reversed''. Thus the simplest SUSY extension of the ISO(4) group
in 4 + 0 dimensions is an $N = 2$ algebra. This algebra when rewritten
in terms of Fermionic creation and annihilation
operators becomes, in the frame where $P^\mu = (0, 0, 0, P)$, 
$$\left\lbrace A_a , A_b^\dagger \right\rbrace = \delta_{ab}
\left[1 + P\left(Z_{Q\overline{Q}}Z^{R\overline{R}}\right)^{-1/2}\right]\eqno(14a)$$
$$\left\lbrace B_a , B_b^\dagger \right\rbrace = \delta_{ab}
\left[1 - P\left(Z_{Q\overline{Q}}Z^{R\overline{R}}\right)^{-1/2}\right].\eqno(14b)$$
We hence see that $P = \sqrt{P_\mu P_\mu}$ has an \underline{upper} bound in $4 + 0$
dimensions if the Hilbert space is to be positive definite,
$$P \leq
\left(Z_{Q\overline{Q}}Z^{R\overline{R}}\right)^{1/2}.\eqno(15)$$
As in 3 + 1 dimensions, saturating the bound eliminates one half of the states.

An important distinction between N = 2 SUSY in Minkowski space and N = 2
SUSY in Euclidean space can now be drawn. In 3 + 1 space the central charge
provides a \underline{lower} bound on the magnitude of the momentum, of
the mass associated with the state. The lower bound can be zero; there is
no inconsistency in considering zero central charge. In 4 + 0 space on
the other hand, the central charge provides an \underline{upper} bound on
the magnitude of the momentum. Such an upper bound on a positive definite quantity
$\sqrt{P_\mu P_\mu}$ cannot be zero; the case of a zero central charge
can only lead to all states having zero momentum yielding a trivial theory.
We conclude that in 4 + 0 space we must include a central charge.

A similar upper bound on momentum arises when one has extended SUSY in $4 + 0$
dimensions with algebra [3]
$$\left\lbrace Q_{ai} , \overline{Q}_{bj} \right\rbrace = \epsilon_{ab}
Z_{ij}^Q\eqno(16a)$$
$$\left\lbrace R_{\dot{a}i} , \overline{R}_{\dot{b}j}\right\rbrace =
\epsilon_{\dot{a}\dot{b}} Z_{ij}^R\eqno(16b)$$
$$\left\lbrace Q_{ai} , \overline{R}_{\dot{b}j}\right\rbrace =
i\sigma_{\dot{a}\dot{b}}^\mu \epsilon_{ij} P^\mu.\eqno(16c)$$

Just as $N = 2$ super Yang-Mills theory in $3 + 1$ dimensions can be
obtained by dimensional reduction of the $N = 1$ gauge theory in $5 + 1$
dimensions, so also the supersymmetric gauge model of Zumino in $4 + 0$
dimensions can be generated; it has the action
$$S = \int d^4 x_E \left[ -\frac{1}{4} F_{\mu\nu}^2 + \frac{1}{2}
\left(D_\mu A\right)^2 - \frac{1}{2} \left( D_\mu B\right)^2 -
\frac{i}{2} \left( \psi^\dagger \gamma \cdot
\stackrel{\leftrightarrow}{D} \psi\right)\right.\nonumber$$
$$\left. + ig\psi^\dagger \left(A - B \gamma_5\right)\psi + \frac{1}{2} g^2 (A
\times B)^2 \right].\eqno(17)$$
One simply drops dependence on one space variable and the time variable in the $5 + 1$
dimensional model and has the corresponding components of the vector field identified
with the scalar fields $A$ and $B$.
Explicit calculation [4] shows that the $\beta$-function in this model
is the same as that in $N = 2$ gauge theory in $3 + 1$ dimensions
despite the peculiar kinetic terms in (16) for the scalars $A$ and $B$.

A model with extended SUSY invariance in $4 + 0$ dimensions can be
obtained by dimensional reduction of $N = 1$ gauge theory in $9 + 1$
dimensions.  It is expected that the $\beta$-function in this model
vanishes, just as it does for $N = 4$ gauge theory in $3 + 1$
dimensions, thereby ensuring that conformal invariance is unbroken.

The $SU(2)$ structure of (12) allows one to define a Harmonic superspace
in conjunction with $4 + 0$ dimensions [5]. This allows for off-shell realization
of this symmetry in these models.

We also note that in $4 + 0$ dimensions, one can define a model which is
(a) Hermitian (b) gauge invariant under an axial $U(1)$ gauge
transformation (c) anomaly free.  Its action is
$$S = \int d^4x_E \left(\frac{1}{4} F_{\mu\nu} (A) F_{\mu\nu}(A) +
\Psi_C^\dagger \left(\not\!p + \not\!\!A \gamma_5 \right) \Psi +
\Psi^\dagger \left(\not\!p - \not\!\!A \gamma_5
\right)\Psi_C\right].\eqno(18)$$
No analogue of this model can be defined in $3 + 1$ dimensions.

The usual form of the actions considered in 4dE are
$$L^{(1)} = \frac{1}{4} F_{\mu\nu} (A) F_{\mu\nu} (A) + \Psi^\dagger
(\not\!\!p + \not\!\!A \gamma_5)\Psi\eqno(19a)$$
or
$$L^{(2)} = \frac{1}{4} F_{\mu\nu} (A) F_{\mu\nu} (A) + \Psi^\dagger
(\not\!\!p + i \not\!\!A \gamma_5)\Psi .\eqno(19b)$$
The former Lagrangian is non-Hermitian while the latter does not have a
compact axial gauge invariance.

\subsection{2 + 2 Dimensions}

In $2 + 2$ dimensions, spinors can be both Majorana and Weyl [6].
Spinors take the form
$$\Psi = \left(\begin{array}{c} \phi_a \\
\chi^{\dot{a}}\end{array}\right)\;\;\;\;
\overline{\Psi} = \left(i \epsilon^{ab} \phi^*_b ,
i\epsilon_{\dot{a}\dot{b}} \chi^{\dot{b}*}\right)\eqno(20a,b)$$
$$\Psi_C = \left(\begin{array}{c} \overline{\phi}_a \\
\overline{\chi}^{\dot{a}}\end{array}\right), \;\;\;\;
\overline{\Psi}_C = \left(\phi^a ,
\chi_{\dot{a}}\right)\nonumber$$
and the two simplest SUSY algebras are
$$\left\lbrace q_a , r_{\dot{b}} \right\rbrace = 2\left(
\sigma_\mu\right)_{a\dot{b}} P^\mu \;\; {\rm{(no\; central \; charge)}}\eqno(21a)$$
and
$$\left\lbrace Q, \overline{Q} \right\rbrace = 2\gamma^\mu P_\mu + Z +
Z_5 \gamma_5 \eqno(21b)$$
for Majorana and Dirac spinorial generators $Q = \displaystyle{\left(\begin{array}{c}
q_a \\
r^{\dot{a}}\end{array}\right)}$ respectively. In [6] it is shown that
both of these algebras can be rewritten in terms of Fermionic creation
and annihilation operators that generate a Hilbert space with negative
norm states; this is taken to indicate that SUSY is incompatible with a
$2 + 2$ dimensional space.

\subsection{3 + 0 Dimensions}
In three dimensional Euclidean space, the simplest SUSY algebra is [7,8]
$$\left\lbrace Q, Q^\dagger \right\rbrace = \vec{\sigma} \cdot \vec{p} +
Z\eqno(22)$$
where $Q$ is a two component Dirac spinorial generator, $\vec{\sigma}$ is a
set of Pauli matrices and $Z$ is a central charge operator.  Forming a
superspace with coordinates $(x^\mu , \zeta , \theta_i$ and $\theta_i^\dagger )$
allows one to make the identifications
$$Q_i = \frac{\partial}{\partial \theta_i^\dagger} - \frac{i}{2}
\left(\vec{\sigma} \cdot \vec{\nabla}\theta\right)_i - \frac{i}{2}
\left(\theta \frac{\partial}{\partial\zeta}\right)_i\eqno(23a)$$
$$P_\mu = -i\frac{\partial}{\partial x^\mu},\;\;\;\;\;
Z = -i\frac{\partial}{\partial\zeta} .\eqno(23b,c)$$

This makes it possible to formulate supersymmetric models in $3 + 0$
dimensions which are analogous to both the Wess-Zumino and $N = 1$ gauge
models in $3 + 1$ dimensions.  A similar analysis can be applied to $N =
2$ supersymmetric models in $2 + 1$ dimensions.  Dimensional reduction
can be used to establish a relationship between supersymmetric models in
$3 + 1$ dimensions and three dimensional supersymmetric models.

We note that just in (8) and (15), in $2 + 1$ dimensions the central
charge in extended SUSY models provides a lower bound for the momentum,  
while in $3 + 0$ dimensions, it provides an upper bound.

An analysis of supersymmetry in five dimensions [7] reveals that much as
in four dimensions, no time dimensions implies an upper bound on
momentum; one time dimension implies a lower bound on momentum and two
time dimensions implies that for all momentum, negative norm states
occur.

\section{Constant Curvature Space SUSY Analysis}
\subsection{$S_2$}
The simplest SUSY algebra [9] associated with the two dimensional
surface of a sphere embedded in three dimensions is
$$\left\lbrace Q_i , Q_j \right\rbrace = 0 \;\; , \;\; \left\lbrace Q_i
, Q_j^\dagger \right\rbrace = Z \delta_{ij} - 2 \vec{\sigma}_{ij} \cdot
\vec{P}\eqno(24)$$
$$\left[J^a, Q\right] = -\frac{1}{2} \sigma^a Q,\;\;\;
\left[J^a, J^b\right] = i\epsilon^{abc} J^c,\;\;\;
\left[Z, Q\right] = -Q\nonumber$$
(Note that $Z$ is no longer a ``central charge'' as it does not commute
with $Q$.).
To examine representations of this superalgebra, we define a state $|I>$
such that
$$\begin{array}{rl}
J^2 |I> \,=\, j(j+1) |I>,& J_3 |I> \,=\, m|I>\\
 & \\
Z |I> \,=\, \zeta |I>, & Q |I> \,=\, 0.\end{array}
\eqno(25)$$
Now if $|i> \,=\, Q_i^\dagger |I>$ and $|F> \,=\, Q_1^\dagger Q_2^\dagger |I>$,
we find that $<1|1> \,=\, (\zeta + 2m)$, $<2|2> \,=\, \zeta - 2m$,
$<F|F> = (\zeta - 2j) (\zeta + 2j + 2)$,
showing that a positive definite Hilbert space occurs if
$$\zeta \geq 2j.\eqno(26)$$
A model invariant under transformations generated by the SUSY algebra of
(22) is
$$S = \int dA \left\lbrace \frac{1}{2} \Psi^\dagger
(\sigma \cdot L + x)\Psi - \Phi^*\left(L^2 + x(1-x)\right)
\Phi - \frac{1}{4} F^*F    \right.\eqno(27)$$
$$\left. + \lambda_N\left(2(1-2x)\Phi^*\Phi - \left(F^*\Phi +
F\Phi^*\right) - \Psi^\dagger \Psi\right)^N\right\rbrace ;\nonumber$$
the off mass shell transformations are
$$\delta\Phi = \xi^\dagger \Psi,\;\; \delta\Psi = 2(\sigma \cdot L+1 -
u)\Phi\xi - F\xi,\; \delta F = -2\xi^\dagger(\sigma \cdot L +
x)\Psi\eqno(28a-c)$$
$$\delta_Z\Phi = \left[ 2(1-2x)\Phi - F\right],\;
\delta_Z \Psi = \left[1 + 2\sigma \cdot L\right]\Psi ,\;
\delta_Z F = -4\left[L^2 + x(1-x)\right]\Phi + 2xF.\eqno(28d-f)$$
The symmetries of (28d-f) are in fact new symmetries.

A superspace representation of the algebra of (22) is provided by
$$Q = (\sigma \cdot \overline{r} + \zeta) \frac{\partial}{\partial
\theta^\dagger} - \left( \frac{\partial}{\partial \zeta} - \sigma \cdot
\nabla\right)\theta\eqno(29a)$$
$$Q^\dagger = \frac{\partial}{\partial
\theta} (\sigma \cdot r + \zeta) + \theta^\dagger
\left( \frac{\partial}{\partial \zeta} - \sigma \cdot
\nabla\right)\eqno(29b)$$
$$J^a = -i(r \times \nabla)^a + \frac{1}{2} \left(\theta^\dagger
\sigma^a \frac{\partial}{\partial \theta^\dagger} +
\frac{\partial}{\partial\theta} \sigma^a\theta \right)\eqno(29c)$$
$$Z = -\theta^\dagger \frac{\partial}{\partial\theta^\dagger} +
\theta\frac{\partial}{\partial\theta}.\eqno(29d)$$

We note that under a supersymmetry transformation generated by (29)
$$\delta r^a = \epsilon^\dagger \sigma^a \theta + \theta^\dagger \sigma^a
\epsilon\eqno(30a)$$
$$\delta\theta = \vec{\sigma} \cdot \vec{r} \epsilon + \zeta\epsilon
.\eqno(30b)$$
Furthermore, we see that
$$\left[ Q, \vec{r}^2 - \zeta^2 - 2\theta^\dagger \theta\right] =
0\eqno(31a)$$
$$\left[ Q, \theta^\dagger \frac{\partial}{\partial \theta^\dagger}
+ \theta \frac{\partial}{\partial \theta} + 
\vec{r} \cdot \vec{\nabla} +  \zeta \frac{\partial}
{\partial\zeta} \right] =
0.\eqno(31b)$$

Currently we are attempting to formulate the model of (25) in terms of superfields
using the superspace realizations (29) of the generators and (31a,b).
\subsection{$AdS_2$}

On $AdS_2$ we have the algebra
$$\left[ J_{ab}, J_{cd} \right] = \eta_{ac} J_{bd} - \eta_{bc} J_{ad} +
\eta_{bd} J_{ac} - \eta_{ad} J_{bc}\eqno(32a)$$
$$\left\lbrace Q, \tilde{Q} \right\rbrace = 2 \Sigma^{ab} J_{ab}\;\;\;\;\;
\left(\tilde{Q} = Q\gamma_2\right)\eqno(32b)$$
$$\left\lbrace J_{ab}, Q\right\rbrace = -\Sigma_{ab} Q\eqno(32c)$$
($Q$ is Majorana, $\eta_{ab} = diag(+, -, +)$, $\gamma_a \gamma_b = -\eta_{ab}
- i\epsilon^{abc}\gamma_c$, $\Sigma_{ab} = \frac{1}{4} \left[\gamma_a , \gamma_b\right]$.)
This algebra can be realized by
$$J_{ab} = \frac{\partial}{\partial \theta} \Sigma_{ab} \theta -
\left(x_a \partial_b - x_b \partial_a\right)\eqno(33a)$$
$$Q = \gamma^a \partial_a \theta + \gamma^a x_a \frac{\partial}{\partial \tilde{\theta}}
\eqno(33b)$$
$$\tilde{Q} = - \tilde{\theta} \gamma^a \partial_a + \frac{\partial}{\partial\theta}
\gamma^ax_a\eqno(33c)$$
where $\theta$ is a two component Grassmann Majorana spinor. We also define
$$D= -\gamma^a \partial_a \theta + \gamma^a x_a \frac{\partial}{\partial\tilde{\theta}}
\eqno(34a)$$
$$\tilde{D} = \tilde{\theta}
\gamma^a \partial_a  +  \frac{\partial}{\partial\theta}
\gamma^a x_a .\eqno(34b)$$
We note that
$$\left[Q_i , x^a \partial_a + \theta_j \frac{\partial}{\partial \theta_
j}\right] = 0\eqno(35a)$$
$$\left[Q_i , x^a x_a - \tilde{\theta} \theta
\right] = 0.\eqno(35b)$$
Applying the condition
$$\Delta \Phi = \omega \Phi\eqno(36)$$
so that if
$$\Phi = \phi + \tilde{\lambda} \theta + F
\tilde{\theta}\theta\eqno(37)$$
then we have $(x \cdot \partial - \omega)$ $\phi = (x \cdot \partial + 1 -
\omega)\lambda = (x \cdot \partial + 2 - \omega)F = 0$.

Some suitable supersymmetric actions are
$$S_1 = \int d^3x d^2\theta \delta\left(x^2 - \tilde{\theta}\theta
- a^2\right)\Phi (\tilde{D}D + \rho)\Phi\eqno(38a)$$
$$S_2 = \int d^3x d^2\theta \delta\left(x^2 - \tilde{\theta}\theta
- a^2\right) \left(\tilde{D}\Phi D\Phi + \rho\Phi^2\right)\eqno(38b)$$
$$S_3 = \int d^3x d^2\theta \delta\left(x^2 - \tilde{\theta}\theta
- a^2\right) \left(\Phi\tilde{Q}Q \Phi + \rho\Phi^2\right)\eqno(38c)$$
$$S_4 = \int d^3x d^2\theta \delta\left(x^2 - \tilde{\theta}\theta
- a^2\right) \left(\tilde{Q}\Phi Q\Phi + \rho\Phi^2\right).\eqno(38d)$$
In component form, for example, (38a) reduces to
$$S_1 = \int d^3x\left\lbrace \delta\left(x^2 - a^2\right)
\left[ F\left(-2x^2 F + 2(\rho-1)\phi\right)\right.\right.\eqno(39)$$
$$\left. + \frac{1}{2x^2} \phi \left(L^{ab} L_{ab} + 2\omega(1 +
\omega)\right)\phi - \tilde{\lambda}\left(\Sigma^{ab} L_{ab}
-\frac{3-\rho}{2}\right)\lambda\right]\nonumber$$
$$\left. + \delta^\prime \left(x ^2 - a^2\right) \left[2\phi\left(x^2 F
+ \left(\frac{\rho}{2} - \omega\right)\phi\right)\right]\right\rbrace
.\nonumber$$
Other supersymmetric actions on $AdS_2$ can be devised in component
form; for example
$$S = \int d^2x \left\lbrace\left[ \tilde{\Psi}\left(\Sigma^{ab} L_{ab}
+ \chi \right) \Phi + \Phi \left( \frac{1}{2} L^{ab} L_{ab} + \chi(1 +
\chi)\right)\Phi\right.\right.\nonumber$$
$$\left.\left.- FF\right] + \lambda_N \left[(1 + 2x)\Phi\Phi +
\tilde{\Psi}\Psi + 2 \Phi F\right]^N\right\rbrace \eqno(40)$$
possesses the invariance
$$\delta \Psi = \left[\left(\Sigma^{ab}L_{ab} - (1 + x)\right)\Phi -
F\right]\xi\eqno(41a)$$
$$\delta\Phi = \tilde{\xi}\Psi\; ,\;\;\;\;\delta F = -
\tilde{\xi}\left(\Sigma^{ab}L_{ab} + \chi \right)\Psi .\eqno(41b,c)$$
The relation between the models of (38) and (40) is not apparent.

The role of $\zeta$ in (29) is not at all clear. However, it is
necessary to introduce $\zeta$ in order for $Q$ to be the
``square root'' of the non-Abelian operator $J^a$.

\begin{center}
{\large\bf{ACKNOWLEDGEMENTS}}
\end{center}
NSERC, Enterprise Ireland--International 
Collaboration Fund 2001 and NUI Galway Millenium Fund 2000 
provided financial support. We would like to thank 
Ochanomizu University for their generous hospitality in organizing 
a most pleasant workshop.  R. and D. MacKenzie had helpful
suggestions\vspace{.7cm}.
\begin{center}
{\large\bf{REFERENCES}}
\end{center}
\begin{itemize}
\item[1.] D. Balin and A. Love, ``Supersymmetric Gauge Field Theory and
String Theory'', IOP Publishing, Bristol 1994\vspace{-.3cm}.
\item[2.] D.G.C. McKeon and T.N. Sherry, Ann. of Phys. 288 (2001) 2\vspace{-.3cm}.
\item[3.] D.G.C. McKeon and T.N. Sherry, Ann. of Phys. 285 (2000) 221\vspace{-.3cm}.
\item[4.] R. Clarkson and D.G.C. McKeon, Can. J. Phys. (to be published)\vspace{-.3cm}.
\item[5.] D.G.C. McKeon, Can. J. Phys. (in press)\vspace{-.3cm}.
\item[6.] F.T. Brandt, D.G.C. McKeon and T.N. Sherry, Mod. Phys. Lett. A 15
(2000) 1349\vspace{-.3cm}.
\item[7.] D.G.C. McKeon, Nucl. Phys. B591 (2000) 591\vspace{-.3cm}.
\item[8.] D.G.C. McKeon and T.N. Sherry, UWO/NUIG report (2001)\vspace{-.3cm}.
\item[9.] D.G.C. McKeon and T.N. Sherry, UWO/NUIG report (2001)\vspace{-.3cm}.
\item[10.] D.G.C. McKeon and T.N. Sherry, UWO/NUIG report (2001).
\end{itemize}
\end{document}